\documentclass[journal,12pt,onecolumn,draftclsnofoot,]{IEEEtran}

\usepackage{amssymb}
\usepackage{flushend}
\usepackage{algorithmic}
\usepackage{array}
\newtheorem{theorem}{Theorem}

\usepackage{pdfpages}
\usepackage{epstopdf}
\usepackage[cmex10]{amsmath}

\usepackage{algorithmic}

\usepackage{array}

\usepackage{pdfpages}

\usepackage{fancyhdr}
\usepackage{kantlipsum}
\renewcommand{\headrulewidth}{0pt}
\renewcommand{\footrulewidth}{0pt}

\fancypagestyle{plain}{
  \fancyhf{}
  \renewcommand{\footrulewidth}{0pt} 
  \renewcommand{\headrulewidth}{0pt}
}
\usepackage{eso-pic}

\let\svthefootnote\thefootnote

\begin{document}

\title{Affinity Scheduling and the Applications on Data Center Scheduling with Data Locality}

\author{
%\IEEEauthorblockN{Amirali Daghighi$^*$}
%\IEEEauthorblockA{Department of Computer Science\\
%University of South Dakota \\ amir.daghighi@coyotes.usd.edu}
%\and
\IEEEauthorblockN{Mohammadamir Kavousi \\}
\IEEEauthorblockA{Department of Electrical Engineering\\
Sharif University of Technology \\ kavousi$\_$mohammadamir@ee.sharif.edu }
}

\maketitle

\thispagestyle{plain}
\pagestyle{plain}

% in the abstract

\begin{abstract}
MapReduce framework is the de facto standard in Hadoop. Considering the data locality in data centers, the load balancing problem of map tasks is a special case of affinity scheduling problem. There is a huge body of work on affinity scheduling, proposing heuristic algorithms which try to increase data locality in data centers like Delay Scheduling and Quincy. However, not enough attention has been put on theoretical guarantees on throughput and delay optimality of such algorithms. In this work, we present and compare different algorithms and discuss their shortcoming and strengths. To the best of our knowledge, most data centers are using static load balancing algorithms which are not efficient in any ways and results in wasting the resources and causing unnecessary delays for users.
\end{abstract}

\IEEEpeerreviewmaketitle

\addtocounter{footnote}{-1}\let\thefootnote\svthefootnote

\section{{INTRODUCTION}}
\label{introduction}

Processing the enormous volume of data sets generated by the social networks such as Facebook
\cite{facebook}
and Twitter
\cite{twitter}
, financial institutions, health-care industry, etc. has become a major motivations for data-parallel processing. In data-parallel processing applications, an incoming task needs a specific piece of data which is physically replicated on different servers. The task receives the service from a server which is close to the physical location of the stored data faster than another server far from the location where the data is stored. For instance, for a task, the speed of receiving the service from a server which has the data stored on its memory or local disk is much faster than receiving the service from a server which does not have the data. That forces the server to fetch the data from another server from the same rack, or even other remote racks. Unless the speed of data center networks have been increased, the differences between service time is still obviously large \cite{intro1}, \cite{intro2}, \cite{intro3}, \cite{nd1}. While assigning tasks, it is a critical consideration to schedule a task on a local server \cite{nd1}, \cite{nd2}, \cite{nd3}, and \cite{nd4}. Scheduling in this setting is called the near-data scheduling problem, or scheduling with data locality. As a result, scheduling and assigning tasks to more suitable servers makes the system stable in the capacity region of the system, and also reduces the mean delay experienced by all tasks in average sense. Therefore, to evaluate different algorithms for scheduling tasks to different servers, two optimality criteria are defined as follows:
\begin{itemize}
\item Throughput Optimality: A throughput optimal algorithm stabilizes the system in the whole capacity region. That is, the algorithm is robust to arrival rate changes, as long as the arrival rate is strictly in the capacity region of the system.
\item Heavy-traffic Optimality: A heavy-traffic optimal algorithm minimizes the mean task completion time as the arrival rate vector approaches the boundary of the capacity region (note that task completion time consists of both waiting time of the task to be assigned to a server and the service time). As a result, a heavy-traffic optimal algorithm assigns tasks to servers efficiently when the system is in the peak loads close to the capacity region boundary. This not only stabilizes the system, but also minimizes the mean delay in stressed load conditions.\\
\end{itemize}

Although there are various heuristic algorithms taking the near-data scheduling into consideration for data centers with multiple levels of data locality \cite{heuristic1}, \cite{nd1}, \cite{heuristic2}, their throughput and heavy-traffic optimality properties are not studied. In this paper, we discuss different real-time scheduling algorithms for systems with multiple levels of data locality with theoretical guaranties for throughput and heavy-traffic optimality. We also compare those algorithms against each other and evaluate them for mean task completion time in the capacity region, including both in high load and low load.

In the following, we first explain the common data center structures and problem definition. In Section \ref{affinity}, we discuss prime algorithms such as Fluid Model Planning and Generalized c$\mu$-Rule. We then discuss the problem for two and then three levels of data locality, in Section \ref{twolevel} and Section \ref{threelevel}, respectively. The paper is concluded in Section \ref{conclusion}.

\section{{SYSTEM MODEL}}
\label{systemmodel}
Consider the system consists of $M$ servers indexed by $m \in \{1, 2, \dots, M \} \overset{\Delta}{=} \mathcal{M}$. Each server belongs to a specific rack denoted by $k \in \{1, 2, \dots, K \} \overset{\Delta}{=} \mathcal{K}$. Without loss of generality, let's assume that the servers in a rack are indexed sequentially. That is, servers $\{ 1, 2, \dots, i \}$ are in the first rack, servers $\{ i + 1, i + 2, \dots, j \}$ are in the second rack, and so on. Let the rack for the server $m$ be $K(m)$. Each data chunk is stored on a set of servers denoted by $\bar{L}$. In real world applications, $\bar{L}$ consists of three servers. The reason to store the data in different servers is to allow the data to be accessed through other servers if a server disconnects from the network or fails to continue working. The larger the set $\bar{L}$ is, the more secure the data would be. However, as the storage of servers is limited, the data is usually stored on no more than three servers. So, $\bar{L} \in \{ (m_1, m_2, m_3) \in \mathcal{M}^3, m_1 < m_2 < m_3 \}$, and the set of all task types is denoted by $\mathcal{L}$. Therefore, different tasks can be labeled by the location of their associated data chunk. For example, all tasks with their stored data in servers $1, 2,$ and $7$ are denoted by type $\bar{L} = \{ 1, 2, 7 \}$ task. All servers in the set $\bar{L}$ are named local servers to the task type $\bar{L}$ since they keep the data needed for the task to be processed. The set of servers $\{ m \notin \bar{L}: \exists n \in \bar{L} \ \text{s.t.} \ K(m) = K(n)  \}$ are rack-local servers, and all other servers are named remote servers to type $\bar{L}$ task. If server $m$ is local, rack-local, or remote to task of type $\bar{L}$, it is denoted by $m \in \bar{L}$, $m \in \bar{L}_k$, or $m \in \bar{L}_r$, respectively. The system model is assumed to be discrete-time. If task $\bar{L}$ is scheduled to server $m \in \bar{L}$ ($\bar{L}_k$, or $\bar{L}_r$), the probability that the task receives service in a time slot and departs the system at the end of the time slot is $\alpha$ ($\beta$, or $\gamma$), where $\alpha > \beta > \gamma$. In other words, the local, rack-local, and remote services follow $Geo(\alpha)$, $Geo(\beta)$, and $Geo(\gamma)$, respectively. As a result, on average it takes less time for a task to receive service from a local server ($\frac{1}{\alpha}$), other than a rack local-local server ($\frac{1}{\beta}$), or a remote server ($\frac{1}{\gamma}$). On the other hand, the arrival of type $\bar{L}$ task at the beginning of time slot $t$ is denoted by $A_{\bar{L}}(t)$ which is bounded with average arrival $\lambda_{\bar{L}}$. The arrival of type $\bar{L}$ tasks is assumed to be i.i.d. \newline
With the service time distribution illustrated above, when a new task arrives, there might not be no local, rack-local, or remote available servers to serve the task immediately. Therefore, multiple queues exist in the data centers where the tasks are kept, waiting to receive service. Based on the structure of the data center and the scheduling algorithm, the number of queues can be less than, equal, or larger than the number of servers. For example, FIFO algorithm just needs one single queue for any number of servers to be implemented. In the rest of the paper, we will point out the number of queues needed for different algorithms. \newline
A question that might be raised here is "At which queue should a new arriving task wait at so to finally receive service from a server?". This part is handled by a \textit{routing} algorithm which takes care of routing new tasks to appropriate queue. On the other hand, when a server is done with processing of a task and becomes idle, it needs to decide which task to pick among all tasks queued at all servers. The act of assigning tasks to idle servers is called \textit{scheduling} with a little bit abuse of terminology. Therefore, an algorithm is fully defined by both its routing and scheduling policies. \newline
As a new terminology, three levels of data locality refers to the case of having all kinds of local, rack-local, and remote services in the system. The number of data locality levels depends directly on the structure of the system. For example, if tasks receive their services just locally or remotely (no rack structure exists), then just two levels of data locality exists.

\subsection{{Capacity Region Realization}}
Let $\lambda_{\bar{L}, m}$ denote the arrival rate of type $\bar{L}$ tasks that receive service from the $m$-th server. In fact, $\lambda_{\bar{L}, m}$ is the decomposition of $\lambda_{\bar{L}}$. Assuming that a server can afford at most load 1 for all local, rack-local, and remote tasks, the capacity region can be characterized as follows \cite{BalancedPandas, yekkehkhany2017near}:

\begin{equation}
\begin{aligned}
\Lambda = & \{  \boldsymbol{\lambda} = (\lambda_{\bar{L}} : \bar{L} \in \mathcal{L}) \ | \ \exists \lambda_{\bar{L}, m} \geq 0, \forall \bar{L} \in \mathcal{L}, \forall m \in \mathcal{M}, s.t. \\
& \lambda_{\bar{L}} = \sum_{m = 1}^{M} \lambda_{\bar{L}, m}, \ \forall \bar{L} \in \mathcal{L}, \\
& \sum_{\bar{L}: m \in \bar{L}} \frac{\lambda_{\bar{L}, m}}{\alpha} + \sum_{\bar{L}: m \in \bar{L}_k} \frac{\lambda_{\bar{L}, m}}{\beta} + \sum_{\bar{L}: m \in \bar{L}_r} \frac{\lambda_{\bar{L}, m}}{\gamma} < 1, \forall m \}
\end{aligned}
\end{equation}

A more thorough look into the definition of the capacity region $\Lambda$, it is easy to figure out that for finding the capacity region of the system described in Section \ref{systemmodel}, a linear programming optimization must be solved.

\section{Affinity Scheduling}
\label{affinity}
The near-data scheduling problem is a special case of affinity scheduling \cite{afs1}, \cite{mandelbaumstolyar}, \cite{intro6}, \cite{afs4}, and \cite{afs5}. In this section, two algorithms, Fluid Model Planning and Generalized c$\mu$-Rule, are illustrated which are somehow the pioneer works on the scheduling problems. However, they are not practical to be used in data centers as it will be discussed in the following two subsections.
\subsection{{FLUID MODEL PLANNING}}
For routing tasks and scheduling servers, fluid model planning algorithm is proposed by Harrison and Lopez \cite{intro5} and \cite{intro6} which is both throughput and heavy-traffic optimal not only for three levels of data locality, but also for the affinity scheduling problem. To implement this algorithm, distinct queues are needed for different types of tasks, $\bar{L}$. Each new incoming task is routed to the queue of its own type. In the model described, there are at most in the order of $M^3$ types of tasks (the data associated to each task type is located on 3 servers, so at most there are ${M}\choose{3}$ $= O(M^3)$ different task types). For finding the scheduling policy, the arrival rate of each task type is required to be known in advance. By solving a linear programming problem, the algorithm distinguishes basic activities. Based on the basic activities derived from the linear programming part, tasks are assigned to idle servers. There are two main objections for this algorithm: First, there should be in the order of $M^3$ number of queues for each task type. In practice, it is not practical to have queues in the cubic order of the number of servers. It excessively complicates the system and its underlying network. Second, the algorithm assumes the arrival rate of different types of tasks to be known. However, firstly the load is not known in real applications, and secondly it changes over time. Therefore, unless the algorithm is throughput and heavy-traffic optimal, it cannot be used in real applications.

\subsection{Generalized c$\mu$-Rule}
Stolyar \cite{stolyar} and Mandelbaum and Stolyar \cite{mandelbaumstolyar} proposed Generalized c$\mu$-Rule. On contrast to fluid model planning which uses the knowledge of the arrival rate of each task type, generalized c$\mu$-rule uses MaxWeight notion which makes the algorithm needless of knowing the arrival rates. But similar to fluid model planning, the algorithm needs one queue per task type. For routing part, each incoming task joins the queue of its own type. Assume that the cost rate incurred by the type $\bar{L}$ tasks queued is $C_{\bar{L}}(Q_{\bar{L}})$, where $Q_{\bar{L}}$ is the queue length of type $\bar{L}$ tasks, and the cost may generally depend on the task type. The cost function should have fairly normal features among which we can mention the followings: $C_{\bar{L}}(.)$ is convex and continuous with $C_{\bar{L}}(0) = 0$, $C^{'}_{\bar{L}}(.)$ to be strictly increasing and continuous with $C^{'}_{\bar{L}}(0) = 0$. Having the cost functions for different kinds of tasks, the server $m \in \mathcal{M}$ is scheduled to a task type $\bar{L}$ in the set below when it becomes idle:

\begin{equation}
\begin{aligned}
\bar{L} \in \underset{\bar{L}}{ArgMax} \ \bigg \{ C^{'}_{\bar{L}} (Q_{\bar{L}}) \mu_{\bar{L}, m} \bigg \}
\end{aligned}
\end{equation}

Where $\mu_{\bar{L}, m}$ is $\alpha$, $\beta$, or $\gamma$ if the task type $\bar{L}$ is local, rack-local, or remote to the idle server $m$ respectively. For instance, if the holding cost for type $\bar{L}$ task is $\gamma_{\bar{L}}Q_{\bar{L}}^{\beta + 1}$ with $\beta > 0$ where satisfies all the conditions for a valid cost function, the generalized c$\mu$-rule is proved by Stolyar to asymptotically minimize the holding cost below \cite{stolyar}:

\begin{equation}
\begin{aligned}
\sum_{\bar{L}} \gamma_{\bar{L}}^{ } Q_{\bar{L}}^{\beta + 1}
\end{aligned}
\end{equation}

As the constant $\beta$ should be strictly positive in order that the cost function satisfies the conditions, the algorithm is not heavy-traffic optimal in the sense defined in Section \ref{introduction}. Besides, using generalized c$\mu$-rule, we still need $O(M^3)$ number of servers (where M is the number of servers). However, it is not practical having a large number of queues as the system becomes more complicated. \newline
All the algorithms given in the next sections do not employ the knowledge of arrival rate, and they assume the system to have the same number of queues as the number of servers, which is a more realistic structure.

\section{{TWO LEVELS OF DATA LOCALITY}}
\label{twolevel}
 The model described in section \ref{systemmodel} is for the case of three levels of data locality, as a task can be served with three different service rates $\alpha$, $\beta$, and $\gamma$. However, most previous theoretical work, except the very last one by Xie, Yekkehkhany, and Lu \cite{BalancedPandas}, has been done on two levels of data locality which is actually the base of three or more levels of data locality. The model for two levels of data locality is somehow the same as the one described in section \ref{systemmodel}; except, in two levels of data locality, there is no notion of rack and rack-local service. Assuming the two levels of data locality, tasks either get service from a server in the set $\bar{L}$ locally with rate $\alpha$, or get service from any other servers remotely with rate $\gamma$. Therefore, the capacity region would be revised as $\sum_{\bar{L}: m \in \bar{L}} \frac{\lambda_{\bar{L}, m}}{\alpha} + \sum_{\bar{L}: m \notin \bar{L}} \frac{\lambda_{\bar{L}, m}}{\gamma} < 1, \forall m \in \mathcal{M}$. For two levels of data locality Wang et al. \cite{MaxWeight}, and Xie, and Lu \cite{Pandas} respectively proposed JSQ-MaxWeight, and Pandas algorithms that are discussed in the next two subsections below.

\subsection{{Join-the-Shortest-Queue-MaxWeight (JSQ-MW)}}
For two levels of data locality, JSQ-MW has been proven to be throughput optimal, but heavy-traffic optimal just in a specific traffic scenario \cite{MaxWeight}. Wang et al. \cite{MaxWeight} assume one queue per server, where the length of queue $m$ at time $t$ is denoted by $Q_m(t)$. A central scheduler maintains the lengths of all queues to decide the routing for new incoming tasks, and scheduling for idle servers. As of routing policy, when a new task of type $\bar{L}$ arrives to the system, the central scheduler routes the task to the shortest queue of the servers in the set $\bar{L}$ (all ties are broken randomly all over this paper). In other words, the new task is routed to the shortest local queue. For scheduling policy, as server $m$ becomes idle, the central scheduler assigns a task queued at a queue in the set below to server $m$:

\begin{equation}
\underset{n \in \mathcal{M}}{ArgMax} \ \{ \alpha Q_n(t) I_{\{ n=m \}}, \beta Q_n(t) I_{\{ n \neq m \}} \}
\end{equation}

Therefore, the idle server gives its next available service to the queue with the maximum weight as defined above. As it was stated before, JSQ-MW is not heavy-traffic optimal in all loads. The specific traffic scenario which the JSQ-MW is heavy-traffic optimal is pointed out in section \ref{evaluation}. For more details refer to \cite{BalancedPandas, yekkehkhany2017near}, and \cite{MaxWeight}.

Next, Pandas algorithm proposed by Xie, and Lu \cite{Pandas} is presented, which is both throughput and heavy-traffic optimal.

\subsection{{Priority Algorithm for Near-Data-Scheduling (Pandas)}}
Pandas algorithm is both throughput optimal and heavy-traffic optimal in all loads \cite{Pandas}. Again, assuming one queue per server, Pandas algorithm routes the new incoming task to the shortest local queue (which is the same as JSQ-MW routing). For scheduling, an idle server always give its next service to a local task queued at its own queue; unless, its queue length is zero. If the idle server's queue does not have any tasks queued, the central scheduler assigns a task from the longest queue in the system to the idle server (which the task is remote to the idle server). This assignment of the remote task from the longest queue in the system occurs when $Q_{max}(t) \geq \frac{\alpha}{\gamma}$, to make sure that the remote task experiences less service time in the remote server other than waiting and receiving its service from the local server.

As the conclusion of the previous work for two levels of data locality, Pandas algorithm proposed by Xie, and Lu \cite{Pandas} is the most promising algorithm that both stabilizes the system in the whole capacity region, and minimizes mean delay for all tasks in heavy-traffic regime. However, in real applications there are usually more than two levels of data locality. The reason is that a server may have the data stored on the memory or local disk, or the server may not have the data saved locally, so it has to fetch the data from another server in the same rack, or even from another server in other racks, which results in appearance of multi levels of data locality. Therefore, it is more of interest to come up with a throughput and heavy-traffic optimal algorithm for a system with more than two levels of data locality. The model illustrated in the system model section has three levels of locality, as a task can get its service locally, rack-locally, or remotely. For the purpose of designing a throughput and heavy-traffic optimal algorithm, assuming three levels of data locality is more challenging than two levels, as a trade-off between throughput optimality and delay optimality emerges. The Pandas algorithm proposed by Xie, and Lu \cite{Pandas} which is both throughput and heavy-traffic optimal for two levels of data locality is not even throughput optimal for three levels of data locality. Xie, Yekkehkhany, and Lu \cite{BalancedPandas} proposed two algorithms for three levels of data locality which are discussed in the next section.

\section{{THREE LEVELS OF DATA LOCALITY}}
\label{threelevel}
For three levels of data locality of which the system model is described in section \ref{systemmodel}, Xie, Yekkehkhany, and Lu \cite{BalancedPandas} extended the JSQ-MaxWeight algorithm and proved it to be throughput optimal. However, the extension of JSQ-MaxWeight is still heavy-traffic optimal only in a specific traffic scenario, not in all loads (note again that JSQ-MW is not also heavy-traffic optimal in all loads for two level of data locality, except a specific load). Xie et al. \cite{BalancedPandas} also proposed a new algorithm called the weighted-workload routing and priority scheduling algorithm which is throughput optimal for any $\alpha > \beta > \gamma$, and is heavy-traffic optimal for the case that $\beta^2 > \alpha \gamma$, which usually holds in real data centers. What $\beta^2 > \alpha \gamma$ implies is that the rack-local service is much faster than remote service. In the next two subsections, JSQ-MaxWeight and the weighted-workload routing and priority scheduling algorithm are discussed in more details.

\subsection{{Extension of JSQ-MaxWeight}}
Assuming one queue per server, the JSQ-MW is as follows: \\
\textbf{Routing:} An arriving task is routed to the shortest queue of the servers in the set $\bar{L}$ (shortest local queue). \\
\textbf{Scheduling:} An idle server is scheduled to a queue in the set
\begin{equation}
\begin{aligned}
\underset{n \in \mathcal{M}}{ArgMax} \ \{ & \alpha Q_n(t) I_{\{ n = m \}}, \beta Q_n(t) I_{\{ K(n) = K(m) \}},\\
&\gamma Q_n(t) I_{\{ K(n) \neq K(m) \}} \}
\end{aligned}
\end{equation}

\begin{theorem}
JSQ-MaxWeight stabilizes the system under any arrival rate vector within $\Lambda$. Therefore, JSQ-MaxWeight is throughput optimal.
\end{theorem}

Proof outline. If $\boldsymbol{\lambda} \in \Lambda$, using $V(t) = \sum_{n = 1}^{M} Q_n^2(t)$ as the Lyapunov function, the $T$ time slot drift of $V(t)$ is bounded in a finite subset of state space of the system, and is negative outside of this subset, which results in stability of the system by Foster-Lyapunov theorem \cite{BalancedPandas, yekkehkhany2017near} (you can refer to \cite{MaxWeight, Pandas, ghassami2017covert} for other uses of Foster-Lyapunov theorem for stability proof of a system).

\begin{theorem}
The extended JSQ-MaxWeight is heavy-traffic optimal in a special scenario of the workload, but not in all traffic scenarios.
\end{theorem}

\subsection{{Weighted Workload Routing and Priority Scheduling}}
Although it is sufficient to have one queue per server to implement the weighted-workload routing and priority scheduling algorithm in a data center with three levels of data locality, it is easier to describe this algorithm assuming the existence of three queues per server. Therefore, assume each server to have three queues, which local, rack-local, and remote tasks to the server are queued in the three queues separately. The central scheduler keeps the vector of queue lengths $\boldsymbol{Q}(t) = (\boldsymbol{Q}_1(t), \boldsymbol{Q}_2(t), \dots, \boldsymbol{Q}_M(t))$ where $\boldsymbol{Q}_m(t) = (Q_m^l(t), Q_m^k(t), Q_m^r(t))$. That is, the first queue of $m$-th server keeps the tasks that are routed to server $m$ and are local to it, the second queue of the $m$-th server keeps the tasks that are routed to this server, but are rack-local to it, and finally remote tasks to the $m$-th server that are routed to this server are queued in the third queue.

Defining the workload on a server, it is ready to give the routing and scheduling policies. The workload on the $m$-th server is defined as the average time needed for the server to give service to all local, rack-local, and remote tasks queued in front of it. The workload on the $m$-th server is defined below:

\begin{equation}
W_m(t) = \frac{Q_m^l(t)}{\alpha} + \frac{Q_m^k(t)}{\beta} + \frac{Q_m^r(t)}{\gamma}
\end{equation}

\textbf{Weighted-Workload Routing:}
As a new task arrives, it joins the server with the least weighted workload. More precisely, the new task joins one of the servers in the following set, where the ties are broken randomly.
\begin{equation}
\underset{m \in \mathcal{M}}{ArgMin} \ \bigg \{ \frac{W_m(t)}{\alpha} I_{\{ m\in \bar{L} \}},\frac{W_m(t)}{\beta} I_{\{ m\in \bar{L}_k \}}, \frac{W_m(t)}{\gamma} I_{\{ m\in \bar{L}_r \}}  \bigg \}
\end{equation}
If the task is local (rack-local, or remote) to the server with the least weighted workload, it joins the first (second, or third) queue which is $Q_m^l$ ($Q_m^k$, or $Q_m^r$).

\textbf{Priority Scheduling:}
When a server, say $m$, becomes idle, it gives its next service to local tasks queued in front of it at $Q_m^l$. In case that there is no local task available for idle server $m$, that is $Q_m^l = 0$, next service is assigned to rack-local tasks queued at $Q_m^k$. Finally if both local, and rack-local queues are empty, the next service goes to $Q_m^r$. In summary, the idle server gives the most priority to local, then rack-local, and finally remote tasks. If all three sub-queues of server $m$ are empty, the server remains idle; until, a new task joins any of the three sub-queues.

\begin{theorem}
The weighted-workload routing and priority scheduling algorithm is throughput optimal, as it stabilizes any arrival rate vector in the capacity region.
\end{theorem}

Proof outline. The $T$ time slot drift of the following Lyapunov function is bounded in a finite subset of state space, and is negative out of this subset, as long as the arrival rate vector is in the capacity region ($\boldsymbol{\lambda} \in \Lambda$), which results in the stability of the system \cite{BalancedPandas, yekkehkhany2017near}.
\begin{equation}
V(t) = ||\boldsymbol{W}||^2 = \sum_{m \in \mathcal{M}} \bigg ( \frac{Q_m^l(t)}{\alpha} + \frac{Q_m^k(t)}{\beta} + \frac{Q_m^r(t)}{\gamma} \bigg )^2
\end{equation}

\begin{theorem}
\label{htobp}
The weighted-workload routing and priority scheduling algorithm is heavy-traffic optimal for $\beta^2 > \alpha \gamma$ \cite{BalancedPandas}.
\end{theorem}

\section{{CONCLUSION}}
\label{conclusion}
In this paper, we first discussed the history of task routing and affinity scheduling in data centers. Two algorithms are then proposed for a system with two levels of data locality: JSQ-MaxWeight, and Pandas algorithms for Near-Data Scheduling (Pandas)- both of which are throughput optimal. However, it was shown that Pandas is the only algorithm being heavy-traffic optimal in all loads. Taking further steps to three levels of data locality, we mentioned that the Pandas algorithm known to be heavy-traffic optimal for two levels of data locality is not even throughput optimal for three levels of data locality. Then, an algorithm with weighted workload routing and priority scheduling as well as an extension of JSQ-MaxWeight are discussed for three levels of data locality. Among these two algorithms only the weighted-workload routing and priority scheduling algorithm is heavy-traffic optimal in all loads.

\bibliographystyle{IEEEtran}
\bibliography{IEEEabrv,qualref}

%\includegraphics[width=1\textwidth]{cv_4.pdf}
%\includepdf[pages=1-]{cv_4.pdf}
\end{document}